\documentclass[10pt,twocolumn,floatfix,superscriptaddress,preprintnumbers,amsmath,amssymb,aps,prb]{revtex4-2}
\usepackage{graphicx}
\usepackage{dcolumn}
\usepackage{bm}
\usepackage{amssymb,amsmath}
\usepackage{float}
\usepackage{color}
\usepackage{ulem}

\begin{document}

\preprint{preprint(\today)}

\title{Pressure-induced superconductivity in monoclinic RhBi$_2$}

\author{KeYuan Ma}
\email{keyuan.ma@cpfs.mpg.de}
\affiliation{Max Planck Institute for Chemical Physics of Solids, Nöthnitzer Straße 40, 01187 Dresden, Germany}
\author{Subhajit Roychowdhury}
\affiliation{Max Planck Institute for Chemical Physics of Solids, Nöthnitzer Straße 40, 01187 Dresden, Germany}
\affiliation{Department of Chemistry, Indian Institute of Science Education and Research Bhopal, Bhopal-462 066, India}
\author{Jonathan Noky}
\affiliation{Max Planck Institute for Chemical Physics of Solids, Nöthnitzer Straße 40, 01187 Dresden, Germany}
\author{Horst Borrmann}
\affiliation{Max Planck Institute for Chemical Physics of Solids, Nöthnitzer Straße 40, 01187 Dresden, Germany}
\author{Walter Schnelle}
\affiliation{Max Planck Institute for Chemical Physics of Solids, Nöthnitzer Straße 40, 01187 Dresden, Germany}
\author{Chandra Shekhar}
\affiliation{Max Planck Institute for Chemical Physics of Solids, Nöthnitzer Straße 40, 01187 Dresden, Germany}
\author{Sergey A. Medvedev}
\affiliation{Max Planck Institute for Chemical Physics of Solids, Nöthnitzer Straße 40, 01187 Dresden, Germany}
\author{Claudia Felser}
\affiliation{Max Planck Institute for Chemical Physics of Solids, Nöthnitzer Straße 40, 01187 Dresden, Germany}

\begin{abstract}
RhBi$_2$ is a polymorphic system that exhibits two distinct phases. RhBi$_2$ in the triclinic phase has been identified as a weak topological insulator with a van Hove singularity point close to the Fermi energy. Thus, triclinic RhBi$_2$ is expected to exhibit exotic quantum properties under strain or pressure. In this study, we report on the emergence of superconductivity in the monoclinic RhBi$_2$ under external pressures. The electrical resistivity behavior of the monoclinic RhBi$_2$ single crystal is studied at a wide range of applied external pressures up to 40 GPa. We observe a pressure-induced superconductivity with a dome-shaped dependence of the critical temperature on pressure at pressures above 10 GPa. A maximum critical temperature ($T_\mathrm{c}$) value of $T_\mathrm{c}$ = 5.1 K is reached at the pressure of 16.1 GPa. Furthermore, we performed detailed ab initio calculations to understand the electronic band structures of monoclinic RhBi$_2$ under varying pressures. The combination of topology and pressure-induced superconductivity in the RhBi$_2$ polymorphic system may provide us with a new promising material platform to investigate topological superconductivity.
\end{abstract}

\maketitle

\section{Introduction}

Superconductivity and topology are two intriguing phenomena of quantum condensed phases in which the former arises due to the electron pairing, and the latter emerges from crystal symmetry and atomic orbital symmetry. These symmetries enrich the characteristics of bands, either as bandgaps or as degeneracies between two or more bands that are protected. Consequently, a fundamental attribute of topological phases is the presence of a gapless state at the boundary. This is related to the topology of the bulk electronic structure and its projection onto the boundary, for example, the surface or the edge \cite{liu2019topological}. The combination of superconductivity and topology leads to another quantum phenomenon -- topological superconductivity, where a topological state and superconductivity couple together in a single material. It is postulated that the such materials host Majorana fermions, which could have potential applications in quantum computing \cite{flensberg2021engineered}. Over the past decades, various topological states, which alone contribute to many exotic electron transports, have been investigated in a range of materials, from insulators to metals, with the aim of understanding their contribution to electron transport. Among these, various Bi-containing topological insulators have been found to be intrinsic superconductors, either by applying chemical pressure via doping \cite{PhysRevLett.104.057001,shruti2015superconductivity,asaba2017rotational} or by applying external pressure \cite{kirshenbaum2013pressure,zhang2011pressure}. However, the origin of superconductivity in these materials remains elusive, varying from one material to other within the framework of bulk and surface superconductivity \cite{sato2017topological,ono2020refined}. Furthermore, the superconductivity observed in the YPtBi topological insulator \cite{butch2011superconductivity,liu2016observation} occurs in the absence of any pressure, and extends beyond the conventional spin-triplet pairing mechanism \cite{kim2018beyond}. Conversely, the Weyl semimetal MoTe$_2$ exhibits superconductivity and displays a significant enhancement under applied external pressure \cite{qi2016superconductivity}. The emergence of superconductivity in these materials not only establishes a connection between superconductivity and topological states, but also suggests the potential for the realization of unconventional superconductivity.

The Bi-rich part of the binary Rh-Bi system has recently attracted research interest due to the superconductivity phenomenon \cite{weitzer2009phase, kainzbauer2018binary}. RhBi$_4$ and Rh$_3$Bi$_{14}$ reveal superconductivity with the transition temperatures of 2.82 and 3.15 K, respectively \cite{weitzer2009phase}. RhBi$_2$ is a polymorphic system that exhibits two distinct phases, each with a unique crystal structure depending on the growth conditions \cite{weitzer2009phase}. The low-temperature $\alpha$-RhBi$_2$ phase has a monoclinic crystal structure, while the high-temperature triclinic $\beta$-RhBi$_2$ phase was realized above 430 $^\circ$C. Both of them are reported to be metallic and non-superconducting at ambient pressure at above 1.8 K \cite{weitzer2009phase,lee2021discovery}. Very recently, the $\beta$-RhBi$_2$ compound with the triclinic phase has been demonstrated to possess week topological insulator properties and is capable of exhibiting intriguing new physical phenomena when subjected to stress or strain \cite{lee2021discovery}.

In this work, we present the results of a study on the electronic transport properties of monoclinic RhBi$_2$ single crystal under high pressures, up to 40 GPa. We show that monoclinic RhBi$_2$ undergoes a pressure-induced superconductivity with a dome-shaped dependence of the critical temperature on pressure at pressures above 10 GPa. Meanwhile, the experimental studies are accompanied by \textit{ab initio} calculations of the phonon dispersion and the electronic band structure under pressures. 

\section{EXPERIMENTAL DETAILS }

Single crystals of RhBi$_2$ sample were synthesized via a Bi flux procedure as reported in previous work \cite{lee2021discovery}. As-purchased high-purity elemental rhodium (99.9\%, ChemPur) and bismuth (99.997\%, Alfa Aesar) were mixed in the molar ratio of Rh:Bi in 1:4. All of the elements were loaded into an alumina crucible which was subsequently vacuum-sealed in a quartz tube under a pressure of $10^{-5}$ mbar. The tube was heated to 1173 K for 12 h, then held for 12 h at this temperature before being progressively cooled to 753 K during 200 h. After centrifuging at 753 K to remove excess Bi, the crystals were recovered. The resulting single crystal has typical dimensions of $4 \times 2 \times 2$ mm$^3$.

Scanning electron microscopy coupled with an energy-dispersive x-ray spectroscopy (EDX) was used to evaluate the composition of the RhBi$_2$ crystal. The single crystallinity of the as-grown crystal was determined by white-beam backscattering Laue X-ray diffraction.

For high-pressure electronic transport studies, a diamond anvil cell equipped with diamond anvils of 500 $\mu$m cutlets was used. For the purpose of electrical resistivity measurements, the tungsten gasket was insulated against the electrical leads with a cubic BN/epoxy mixture. A single crystalline sample of a near-square shape with $\sim 120$ $\mu$m size and $\sim 10$ $\mu$m thickness was loaded into the sample chamber (diameter 200 $\mu$m and height 40 $\mu$m) ﬁlled with NaCl as the pressure transmitting medium. Four electrical leads, fabricated from 5-$\mu$m-thick Pt foil, were contacted to the surface of the sample in the van der Pauw conﬁguration. The electrical resistivity was measured at different pressures on a Quantum Design Physical Property Measurement System (PPMS) using low-frequency ac excitation at temperatures down to 1.8 K. The pressure was determined using the ruby luminescence method.

In order to perform the ab-initio calculations of the electronic structure, we employ the density-functional theory, which is implemented in the VASP software \cite{kresse1993ab,kresse1994ab,kresse1996efficiency,kresse1996efficient}. The exchange-correlation potential is approximated by the generalized gradient approximation as parametrized by PBE \cite{perdew1996generalized}. The influence of external pressures was accounted for using the PSTRESS functionality of VASP. For each pressure value, the crystal structure was relaxed until all forces were below 10$^{-4}$ eV/Ang. For the phonon calculations, the PHONOPY package \cite{togo2015first} was employed together with VASP and a 2 x 2 x 2 supercell.

\section{RESULTS and DISCUSSION}

\begin{figure}[htb!]
\includegraphics[width=1.0\linewidth]{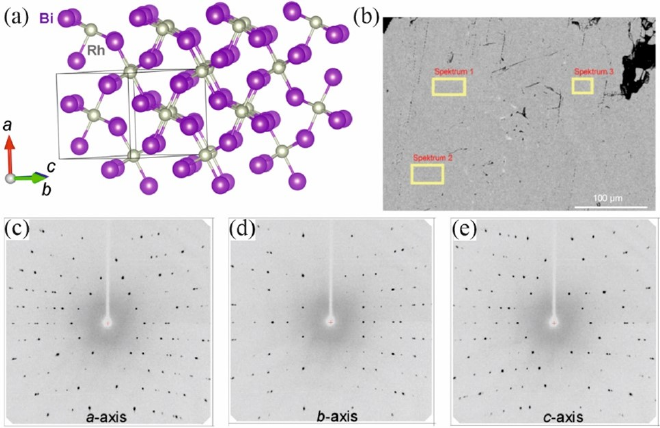}
\caption{Structural characterization of the grown single crystal of monoclinic RhBi$_2$: (a) Crystal structure of monoclinic RhBi$_2$. (b) Back-scattered scanning electron microscopy image of a slice of monoclinic RhBi$_2$ wherein labeled regions show the regions taken into consideration for EDX elemental composition analysis; (c), (d), (e) Selected oscillation images of a monoclinic RhBi$_2$ single crystal along the $a$, $b$, and $c$ axes, respectively.}
\label{fig1}
\end{figure}

Single crystal X-ray diffraction oscillation patterns confirm the high-quality of the obtained single crystal of RhBi$_2$. The stoichiometry of the crystals was found to be close to the nominal composition and uniform across the sample, as revealed by an EDX analysis (Fig.\ 1, Table 1). The monoclinic unit cell of $P21/c$ symmetry with lattice parameters of $a$ = 6.9241(13) \AA, $b$ = 6.7946(8) \AA, and $c$ = 6.9587(13) \AA, $\alpha$ = $\beta$ = 90$^\circ$, $\gamma$ = 117.73$^\circ$ (Fig. 1) were determined by unambiguous indexing of the x-ray diffraction patterns, which is in full agreement with the published data for the stable monoclinic $\alpha$-phase of RhBi$_2$ at room temperature \cite{weitzer2009phase,kjekshus1971properties}. In contrast to the high-temperature triclinic $\beta$-phase modification \cite{weitzer2009phase,ruck1996kristallstruktur}, the monoclinic structure lacks van der Waals-like bonding along the crystallographic $a$ axis.

\begin{table} [H]
\centering
\caption{EDXS analysis for elements Rh and Bi.}
\begin{tabular}{|c|c|c|}
\hline
Region (Fig.1(b)) & Rh element (\%) & Bi element (\%) \\ \hline
1                 & 34.69           & 63.23           \\
2                 & 34.62           & 64.02           \\
3                 & 35.91           & 62.80           \\ \hline
Average           & 35.07           & 63.68           \\ \hline
\end{tabular}
\label{table:Table1}	
\end{table} 

At ambient pressure, the electrical resistivity of the monoclinic RhBi$_2$ decreases with decreasing temperature, revealing a metallic behavior. To see the effect of pressure, we measured the temperature dependence of the resistivity of the monoclinic RhBi$_2$ at different pressures, as shown in Fig.\ 2a. At pressures up to 10.6 GPa, the resistivity decreases upon cooling at all pressures, indicating that monoclinic RhBi$_2$ exhibits a normal metallic behavior. No evidence of superconductivity is observed upon cooling down to 1.8 K in this pressure range. Notably, the elemental bismuth is reported to become superconducting at temperatures below 4-8 K in a pressure range of 2.6–10.5 GPa \cite{li2017pressure}. The absence of a superconducting transition in our resistivity measurements at low pressures reveals the absence of Bi-impurity in the sample under our study. 

As pressure is increased up to 10.6 GPa, the abrupt drop in the resistivity at temperature below 2 K indicates the possible onset of a superconducting transition (Fig.\ 2a, b). Indeed, as pressure increased up to 13.2 GPa the zero resistivity can be observed (Fig.\ 2b), indicating the transition of the monoclinic RhBi$_2$ to the superconducting state. 

\begin{figure}[htb!]
\centering
\includegraphics[width=1.0\linewidth]{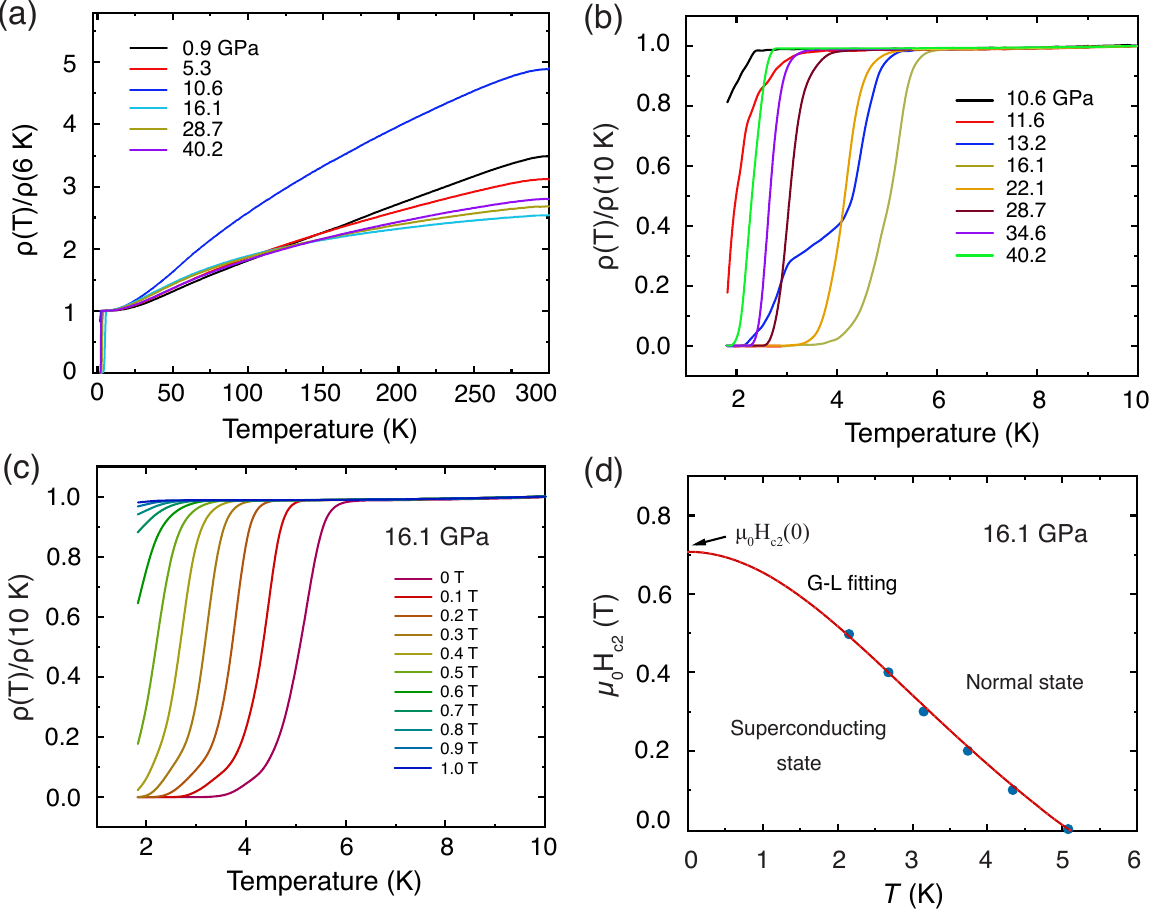}
\caption{(a) Temperature dependence of the normalized electrical resistivity of the monoclinic RhBi$_2$ under different pressures. (b) Temperature dependence of the normalized resistivity in the vicinity of the superconducting transition. (c) Temperature dependence of the normalized resistivity under different magnetic fields at 16.1 GPa. (d) Temperature dependence of upper critical field for monoclinic RhBi$_2$. Here, $T_\mathrm{c}$ is determined as the 50\% drop of the normal state resistivity. The solid line represents the fit based on the Ginzburg–Landau (GL) formula.}
\label{fig2}
\end{figure}

The superconducting state is further confirmed by the evolution of the temperature dependence of the resistivity in an applied magnetic field, as shown in Fig.\ 2c. The superconducting transition gradually shifts toward lower temperatures with the increase of the magnetic ﬁeld. At the applied field of $\mu_0 H$=1 T, all signs of superconductivity are disappeared at temperature above 1.8 K. Fitting of the critical field data $\mu_0 H_\mathrm{c2}(T)$ to the Ginzburg-Landau equation \cite{tinkham1996introduction}:
\begin{equation}
\mu_0 H_\mathrm{c2}(T) = {\mu_0 H_\mathrm{c2}(0)} \frac {(1-t^2)}{(1+t^2)}.
\label{G-L fitting}
\end{equation}
where $t = T/T_\mathrm{c}$ is the reduced temperature and $T_\mathrm{c}$ is the transition temperature at zero magnetic field, yields the upper critical field value $\mu_0 H_\mathrm{c2}(0)$ = 0.71 T for the monoclinic RhBi$_2$ at 16.1 GPa (Fig. 2d). The obtained value of the critical field yields (according to the relationship: \begin{equation}
\mu_0H_\mathrm{c2}(0) = \frac{\Phi_0}{2 \pi \ \xi_\mathrm{GL}^2}.
\label{eq:GL}
\end{equation}
where $\Phi_0 = h/(2e) \approx 2.0678 \times 10^{-15}$ Wb is the quantum flux) a Ginzburg–Landau coherence length of $\xi_\mathrm{GL} = 21.5$ \AA.  The obtained value of the upper critical field $\mu_0H_\mathrm{c2}(0)$ is significantly below the Pauli limit. 

In the absence of the experimental data on the pressure evolution of the structure of RhBi$_2$, the dynamical stability of monoclinic RhBi$_2$ under compression was carefully examined by calculating the phonon spectrum of the monoclinic RhBi$_2$ structure at different pressures up to 40 GPa (Fig. 3). It can be seen that no imaginary phonon modes are revealed by the calculations, indicating that the monoclinic RhBi$_2$ structure is dynamically stable within the studied pressure range. The phonon band structure calculations show a progressive hardening of the acoustical modes with increasing levels of compression, which also indicates the structural stability of the compound over the full pressure range. 

\begin{figure}[htb!]
\centering
\includegraphics[width=0.85\linewidth]{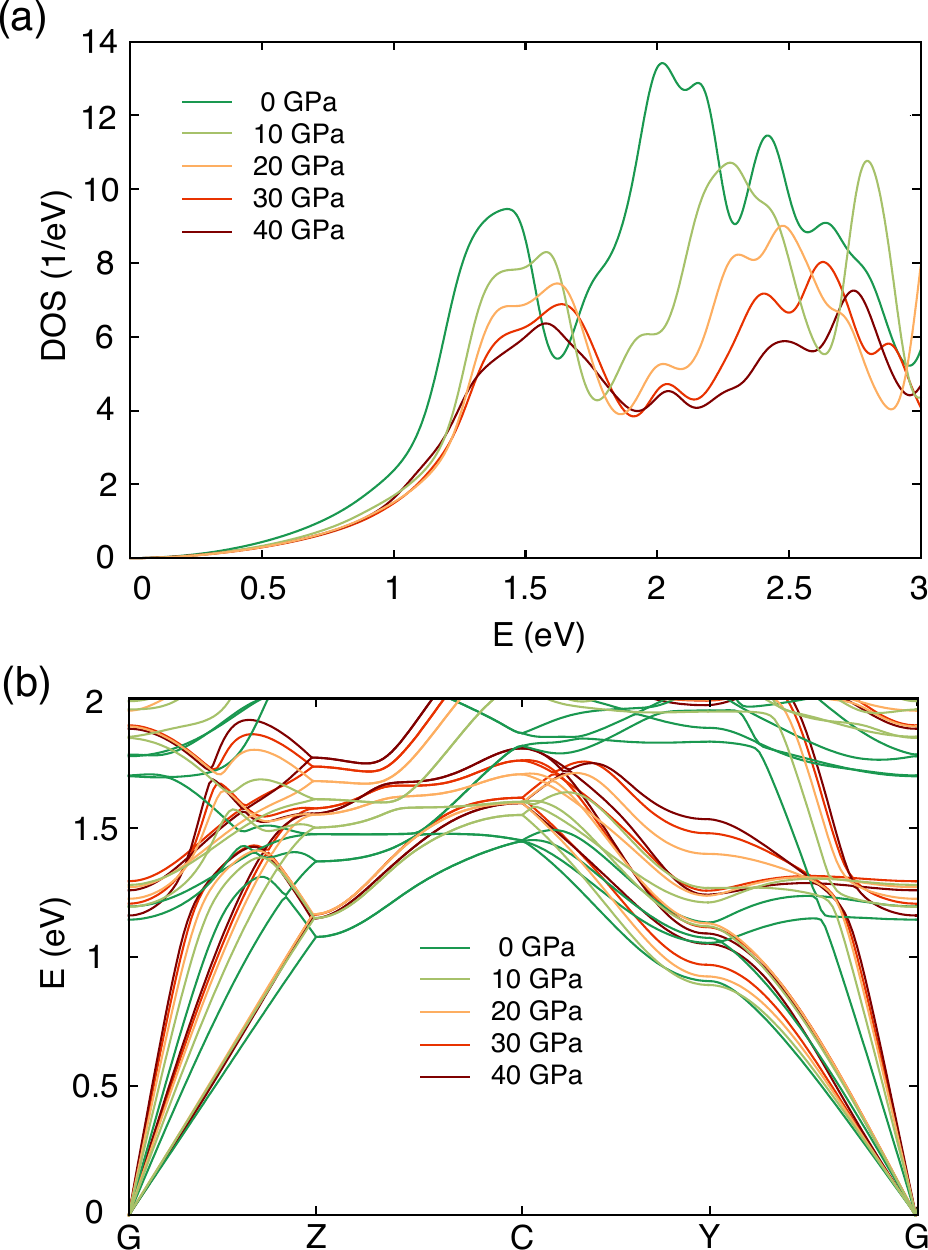}
\caption{Phonon dispersion of the monoclinic RhBi$_2$ under pressure. (a) Phonon density of states (DOS) for different pressure values. (b) Phonon band structure for different pressure values. Dynamical stability for the whole pressure range is implied.}
\label{fig3}
\end{figure}

\begin{figure}[htb!]
\centering
\includegraphics[width=1.0\linewidth]{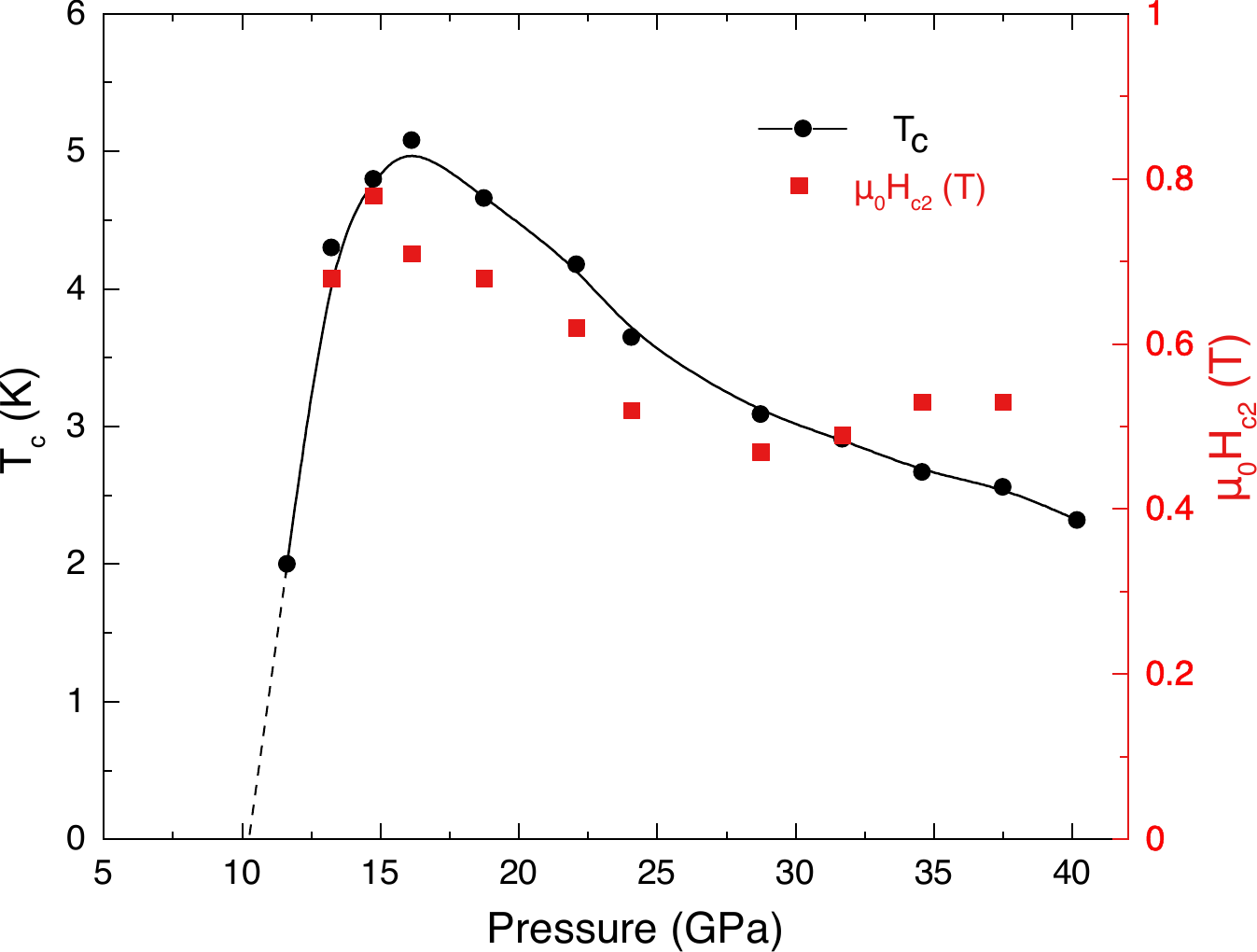}
\caption{Pressure dependence of the critical temperature of superconductivity $T_\mathrm{c}$ and upper critical field $\mu_0H_\mathrm{c2}(0)$ in monoclinic RhBi$_2$. $T_\mathrm{c}$ is determined as the 50\% drop of the normal-state resistivity.}
\label{fig4}
\end{figure}

The pressure evolution of the superconducting transition in monoclinic RhBi$_2$ was further studied by measuring the resistivity at pressures up to above 40 GPa (Fig.\ 2b). The superconducting transition exhibits a rapid shift towards higher temperatures as the pressure increases up to values above 16 GPa. The maximum value of the $T_\mathrm{c} = 5.1$ K is reached at pressure 16.1 GPa. As the pressure is increased further, the $T_\mathrm{c}$ decreases gradually, reaching a minimum of 2.3 K at the highest pressure in this experiment, 40.2 GPa. This results in a dome-shaped dependence of the $T_c$ on pressure, as shown in Fig. 4. As anticipated, the value of the upper critical field $\mu_0H_{\rm c2}(0)$ decreases gradually with pressure increase down to 0.53 T at 37.5 GPa in correlation with the decrease of the $T_\mathrm{c}$ (Fig.\ 4). At all pressures, $\mu_0H_{\rm c2}(0)$ value remains below the Pauli limit, a phenomenon observed in numerous metal-Bi binary superconducting alloys, including KBi$_2$, RbBi$_2$, and $\beta$-PdBi$_2$ \cite{li2020new,li2022pressure,imai2012superconductivity,sakano2015topologically}.

\begin{figure}[htb!]
\centering
\includegraphics[width=0.9\linewidth]{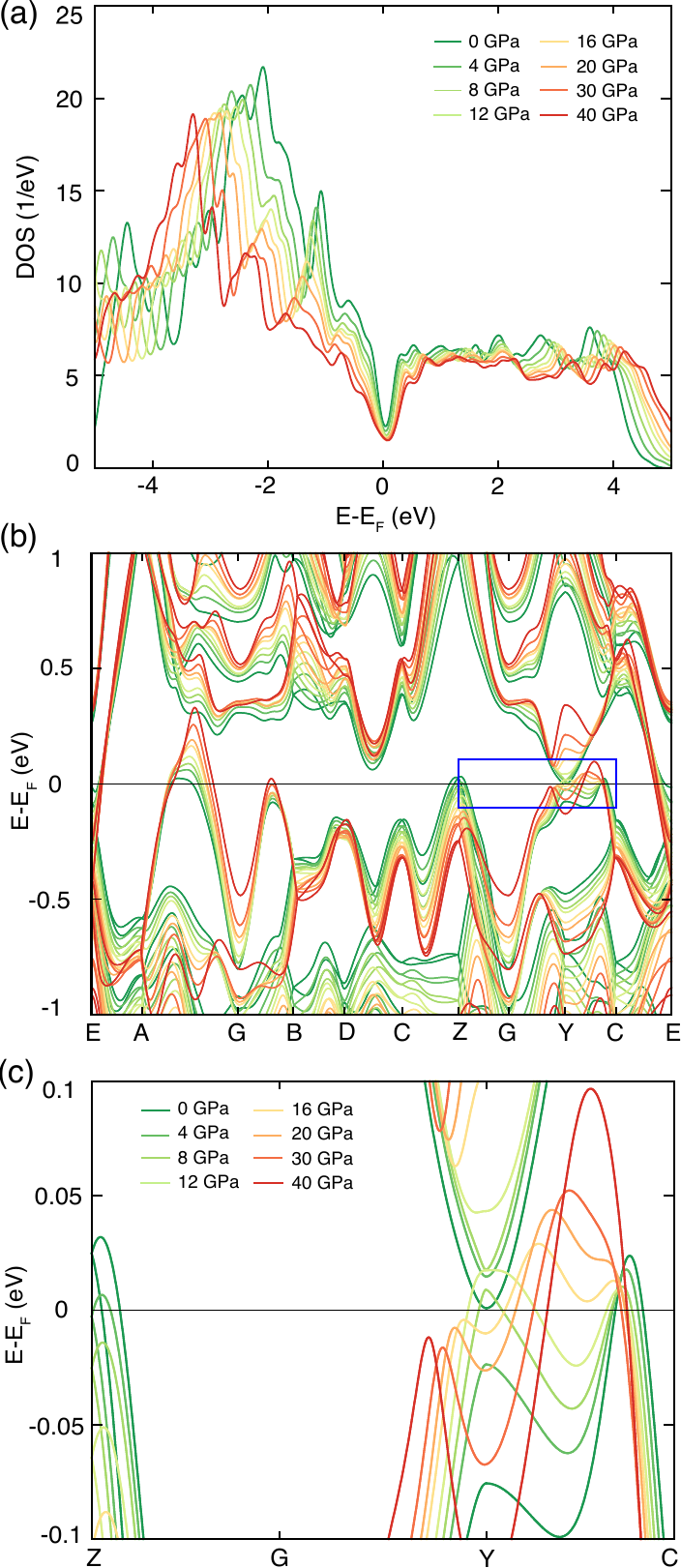}
\caption{Electronic structure of monoclinic RhBi$_2$ under pressure. (a) Electronic density of states (DOS) for different pressure values. (b) Electronic band structures for different pressure values. (c) Zoom-in of the electronic bands to the blue rectangle in (b), showing the move of the valence band at Y through the Fermi energy twice with increasing pressure.}
\label{fig5}
\end{figure}

To get further insight into the pressure effect on the electronic properties of monoclinic RhBi$_2$, its electronic band structure was calculated at different pressures. In Fig.\ 5a, the pressure evolution of the electronic density of states (DOS) is shown. It can be seen that the pressure has a minor effect on the overall shape of the DOS. The only noticeable effect is a stretching of the energy with increasing pressure, indicating a reduction of the DOS at and at the vicinity of the Fermi energy. For a more detailed analysis, the electronic band structure is also investigated and shown in Fig.\ 5b. The increase in pressure leads to an increase in band energies in the conduction band everywhere except the narrow k-region around the E point, while for the valence bands no systematic behaviour can be determined. Additionally, a very interesting feature can be found around the Y point (see Fig.\ 5c). Here, the valence band is first rising in energy with pressure and at  pressure 3 GPa the band crosses the Fermi energy and the band energy is further increasing as the pressure increases up to the 12 GPa. At this pressure the band energy at the Y point reaches its maximum and starts to get lower in energy, as the pressure cntinue to increase. At pressure 14 GPa, the band crosses the Fermi energy again and remains below the energy level up to the pressure 40 GPa. Such non-monotonous  behavior resembles remarkably the dome-shaped dependence of $T_\mathrm{c}$ on pressure. At ambient pressure, RhBi$_2$ is a trivial insulator. With increasing pressure, around the Y point a band inversion emerges above 14 GPa. However, this does not change the system into a topological insulator, because at the same time, an enforced degeneracy at the U line also emerges. This leaves the high-pressure state of RhBi$_2$ as an enforced semimetal.

\section{Conclusion}
In summary, we have demonstrated a high-pressure study of the electrical resistivity behavior of the monoclinic RhBi$_2$ accompanied by theoretical calculations of the electronic band structures. The monoclinic RhBi$_2$ is non-superconducting under ambient pressure, however, it shows superconductivity under external pressures at above 10 GPa. At the measured pressure range up to 40 GPa, the critical temperature ($T_\mathrm{c}$) of superconductivity demonstrates a dome-shaped dependence on pressures. The maximum critical temperature value of $T_\mathrm{c} = 5.1$ K is observed at the pressure of 16.1 GPa. The theoretical calculations of the phonon spectrum at different pressures indicate the dynamical stability of the monoclinic RhBi$_2$ structure over the whole studied pressure range. The observed pressure induced superconductivity in the monoclinic RhBi$_2$ may provide a new promising platform to study topological superconductivity.

\begin{acknowledgments}
This work is financially support by the Deutsche Forschungsgemeinschaft (DFG) under SFB1143 (Project No.\ 247310070), the Wuerzburg-Dresden Cluster of Excellence on Complexity and Topology in Quantum Matter–ct.qmat (EXC 2147, Project No.\ 390858490)
\end{acknowledgments}

\end{document}